\documentclass[letterpaper,twocolumn,10pt]{article}
\usepackage{usenix-2020-09}

\usepackage{tikz}
\usepackage{amsmath}
\usepackage{hyperref}
\usepackage{cleveref}
\usepackage{enumitem}
\setlist[itemize]{leftmargin=*,noitemsep,topsep=0pt,parsep=0pt,partopsep=0pt}
\setlist[enumerate]{leftmargin=*,itemsep=20pt,topsep=0pt,partopsep=0pt}
\usepackage{lipsum}
\usepackage{caption}
\usepackage{subcaption}
\usepackage{graphicx}
\usepackage{layout}
\newif\ifdbg
\dbgfalse 

\newif\iftemplate
\templatetrue

\newif\ifhidden
\hiddenfalse

\begin{document}

\date{}

\title{\Large \bf CheckSync: Using Runtime-Integrated Checkpoints to Achieve High Availability}

\author{
    {\rm Nicolaas Kaashoek}\\
    Princeton University
    \and
    {\rm Robert Morris}\\
    MIT CSAIL
}

\maketitle
\begin{abstract}
CheckSync provides applications with high availability via runtime-integrated checkpointing. This allows CheckSync to take checkpoints of a process running in a memory-managed language (Go, for now), which can be resumed on another machine after a failure. CheckSync uses the runtime to checkpoint only the process' live memory, doing without requiring significant changes to applications.

CheckSync maintains the ease of use provided by virtual machines for the applications it supports without requiring that an entire virtual machine image be snapshotted. Because CheckSync captures only the memory used by an application, it produces checkpoints that are smaller (by an order of magnitude) than virtual machine snapshots if the memory footprint of the application is relatively small compared to the state of the rest of the operating system. Additionally, when running go-cache, a popular in-memory key/value store, CheckSync reduces throughput by only 12\% compared to the 78\% throughput loss when using \texttt{go-cache}'s snapshot functionality, the 45\% loss when using CRIU, and the 68\% loss when using virtual machine live migration.
\end{abstract}
\section{Introduction}
\label{sec:intro}
Having a general-purpose way of providing high availability is important, especially to critical components of distributed systems like the MapReduce coordinator\cite{dean_mapreduce_2008} or lock servers\cite{burrows_chubby_2006}. Adding component-specific solutions to these components accomplishes the goal of making them highly available, but adds extra complexity to simple pieces of code. Providing an easy-to-use interface that works for many applications is therefore better. 

One commonly used solution is to capture the state of the application in its entirety and replicate that state. This is agnostic to the application running, which makes it a particularly easy solution to  fault tolerance to a system. Often, this is done using virtual machines where the hypervisor takes a snapshot of the running virtual machine that the system can then resume from\cite{setty_vmware_2021,bressoud_hypervisor-based_1995,cui_snapshot_2015}. This includes data extraneous to the application, such as state needed to reconstruct the virtual machine and OS kernel.

This idea is extended in virtual machine live migration\cite{clark_live_2005,cully_remus_2008,setty_vmware_2012}. At a high level, a primary hypervisor copies the snapshots to a backup location, which then resumes and takes over if the primary fails. This extends the fault tolerance provided by snapshots with high availability, but still requires that extra state be replicated.

This paper presents CheckSync, the first system to provide high availability by consulting memory information maintained by the language runtime to construct checkpoints. This maintains the ease of use of virtual machines for a restricted class of applications, while providing an increase in performance. CheckSync periodically checkpoints while an application executes, dumping the essential state of the application to network storage as a file. A backup machine monitors the primary to detect failover.

After detecting a failure, the backup fetches the checkpoints from network storage and loads them into memory to resume application execution. To any clients interacting with the system, this looks like a network error, and upon retrying they are directed to the backup and can resume execution. CheckSync relies on a Paxos-replicated configuration management service to handle the forwarding and assignment of primary and backup machines.

Checkpointing applications for high availability presents three challenges. First, CheckSync must manage application multithreading to ensure that it is in a state where it can be checkpointed. Second, CheckSync must keep the size of checkpoints small in order to be performant. The larger a checkpoint is, the longer CheckSync has to spend writing to disk, during which time application performance is paused. Third, CheckSync must seamlessly resume from the checkpoint after a failover; this means it has to construct a process on the backup that is identical in memory, file descriptors, and registers as the checkpointed process on the primary. Doing so is tricky.

CheckSync is designed specifically to support applications written in memory-managed languages. The runtimes of these languages track and update information about the state of the running program, especially in the garbage collector. CheckSync uses this runtime state to safely pause the execution of the application and to determine what pages need to be dumped in each checkpoint.

CheckSync's access to the runtime allows it to operate without relying on the application to tell it what to replicate and when. Instead, the application runs on the modified runtime and CheckSync will determine where in memory the application lives and what needs to be check pointed. It uses the runtime and garbage collector to ensure that checkpoints happen only at times that are safe for the application, such as ensuring that no thread is holding any locks in the runtime, which could otherwise lead to deadlock. This makes CheckSync easy to use for application developers.

CheckSync only checkpoints memory that the runtime indicates is live, not memory that is on the free list. Compared to virtual machines, which have to snapshot all of the operating system's memory, CheckSync is able to use this technique to produce smaller checkpoints. This improves performance, as the smaller the checkpoint is, the quicker it is to take as less disk writes are required. 

Replication in CheckSync is asynchronous by default, as the application is paused only while the checkpoint is taken, and resumes before the checkpoint is persisted anywhere. This means that clients may receive responses to requests that aren't included in the latest checkpoint. These will then be lost after a failover. However, CheckSync provides a synchronous operating mode as well, where the application's execution is resumed only after the checkpoint is successfully replicated. Application developers must mark where in the code they return a response to clients, and CheckSync will then take a checkpoint at these locations. This allows CheckSync to ensure that all state made visible to clients is included in the latest checkpoint.

Runtime integration does restrict the kinds of applications that CheckSync can support. Because CheckSync lives in the same process as the application, it cannot replicate ones that rely on multiple processes or any form of inter-process communication, or any state that is stored in the kernel such as TCP connection state. This means that applications that use CheckSync have to reconnect after a failover rather than having their connections carried over. Additionally, applications are responsible for handling duplicate request detection, as that may arise when the clients reconnect after failover. Finally, while CheckSync replicates open file descriptors, its file transferal is rudimentary and performance suffers if large files are replicated. Applications that cannot handle the loss of data that may occur after a failover in asynchronous CheckSync are forced to use synchronous CheckSync.

We implemented CheckSync for Go. We evaluate its ease of use with three sample Go applications: the \texttt{go-cache} key/value store\cite{nielsen_go-cache_2021}, a coordinator for a MapReduce-like system, and a mathematical benchmark\cite{noauthor_github_nodate} that simulates a long-running compute job. The evaluation demonstrates that CheckSync is easy-to-use, requiring only 5 lines of code to be changed for the MapReduce coordinator, and none for the other applications. 

We also evaluate CheckSync's performance, showing that it produces checkpoints significantly smaller than virtual machine snapshot, while reducing throughput by only 12\% when checkpointing ever 200ms. \texttt{Go-cache}'s application-specific snapshots are smaller than CheckSync's checkpoints, but they reduce throughput by 78\%. Remus\cite{cully_remus_2008} imposes a 68\% loss of throughput. CRIU\cite{noauthor_checkpoint-restorecriu_2021}, a process checkpointer, produces checkpoints 3x larger than CheckSync at a 45\% loss. CheckSync with synchronous replication has a significantly higher overhead.

CheckSync's main contribution is its exploitation of language runtime state to reduce checkpoint cost. This makes CheckSync an easy-to-use, high-performance solution to providing mission-critical applications with high availability. The rest of this paper presents CheckSync's design, its implementation, and an evaluation of its performance. 
\section{Related Work}
\label{sec:related}
This section reviews categories of related work roughly going from more invasive approaches that require more work on the part of application developers to less invasive approaches.

\textbf{Replicated State Machines}. Replicated state machines\cite{schneider_implementing_1990} are widely used to provide high availability. Paxos\cite{lamport_paxos_2001} and the family of algorithms it spawned are widely used today, as is Raft\cite{ongaro_search_2014}. Indeed, CheckSync relies on the presence of a consensus protocol to handle configuration management, as these protocols can handle network partitions.

However, adapting existing applications to use these services can be difficult\cite{chandra_paxos_2007,bolosky_paxos_2011}, as not all applications can be implemented as state machines without significant redesign. For example, consider the MapReduce coordinator, which tracks a set of tasks and distributes them to workers. To use state machine replication, the coordinator would have to be redesigned to store the set of tasks in a log, and to remove them from the log in response to worker activity. This requires a significant effort on the part of the application designer. 

Additionally, state machine replication prevents applications from using mulithreading as they require deterministic execution. This limits the performance of applications that use these systems by removing the benefits of parallelism. While attempts have been made to address this weakness\cite{guo_rex_2014,eve_2012}, these systems still require applications to be redesigned to fit their protocols, making them difficult to use.

Another way to address the above problem is with deterministic execution\cite{altekar_odr_2009,cui_efficient_2011,basile_active_2006}. When an application executes deterministically, it can use replicated state machines\cite{paxos_transparent} even with multithreading. While deterministic execution is an effective technique for enabling state machine replication in multithreaded applications, it adds a sizable overhead to application performance\cite{bergan_deterministic_2011}.

\textbf{Checkpointing and Snapshotting}. CheckSync is not the first system to use checkpointing for fault tolerance. Many database systems\cite{noauthor_creating_nodate,ren_low-overhead_2016} also use checkpoints to recover from crashes. Some of these systems, including members of the SQL family\cite{postgres_replication,litestream}, provide services to replicate these checkpoints as well. Distributed storage systems like Amazon Aurora\cite{amazon_aurora}, Amazon RDB\cite{noauthor_creating_nodate}, Ceph\cite{weil_ceph_2006}, and NFS\cite{sandberg_sun_1986} can also snapshot and replicate their persistent state to increase availability.

All of these systems snapshot only the persistent data stored in the system, such as the file system in Ceph or the database in PostgreSQL. This makes them excellent solutions for applications that only care about making their data highly available. However, CheckSync is also capable of replicating and restoring the computation state of the application. This allows it to support a class of applications that the above systems cannot.

\textbf{Process-Level Checkpointing}. Other systems can checkpoint processes as well, such as DMTCP\cite{ansel_dmtcp_2009,noauthor_dmtcp_2021}, CRIU\cite{noauthor_checkpoint-restorecriu_2021} and FTI\cite{leo_what_2021}. DMTCP and CRIU are widely used today, for scientific computing and container snapshotting respectively. These frameworks are able to replicate state that CheckSync cannot, such as open X11 windows and the full TCP stack. While this allows them to support applications that CheckSync can't, they do not have access to the runtime integration that CheckSync does. With the information it gets from the runtime, as well as the smaller amount of state that needs to be replicated, CheckSync producese checkpoints that are smaller than those from these process checkpointing services (3x smaller than CRIU, for example - \cref{tab:checkpointsize}).

Checkpoint services have also been designed to support serverless computing. The use case and motivation behind these systems is different than for CheckSync. They focus on reducing start-up costs for serverless functions\cite{zhang_kappa_2020,du_catalyzer_2020}, rather than providing high availability. They are optimized for the restore process, while CheckSync is more concerned with keeping the checkpointing as fast as possible. Additionally, many of them lack support for multithreaded applications as serverless functions are generally single-threaded.

\textbf{Virtual Machine Live Migration}. Virtual machines have long been seen as a potential technique for achieving transparent fault tolerance using snapshots\cite{cui_snapshot_2015,setty_vmware_2021}. They have also been used to provide high availability using live migration\cite{clark_live_2005}. VMWare proposed one such method for doing so using lockstep replication in VMWareFT\cite{scales_design_2010}. That solution has evolved over time into vSphere's vMotion replication option\cite{setty_vmware_2012}, an asynchronous technique for disaster recovery. Open source hypervisors like Xen\cite{xen_hypervisor} provide similar tools through the use of Remus\cite{cully_remus_2008}. These tools are extremely easy to use, and almost any application can run on them. However, this comes at the tradeoff of having larger checkpoints, as virtual machines snapshots must include the state of the entire operating system. CheckSync instead sacrifices some general-purpose ease of use to reduce the amount of state that is checkpointed, which reduces its overhead.

\textbf{Containerization}. One alternative to using a virtual machine is to run applications inside a container such as Docker\cite{merkel2014docker} or LXC\cite{LXC}, which also have snapshot capabilities. While this provides the same ease of use as virtual machine snapshotting with a lower overhead due to containers being more lightweight, there is no live migration solution for container-based systems\cite{torre_towards_2019}. Like virtual machines, containers are easier to use than CheckSync and support more applications, but have a higher overhead.

\section{Design}
\label{sec:design}
CheckSync makes applications highly available by first \emph{checkpointing} the application, and then \emph{synchronizing} the checkpoint with a backup that can then restore from that checkpoint in the face of failure. The design aims to minimize the cost of checkpointing on application performance by exploiting state in the language runtime to reduce the size of the checkpoints that it takes.

CheckSync is designed for applications that live in-memory and are critical to the operation of a larger system. Three examples of such applications are the MapReduce coordinator, configuration managers, and long-running, compute applications. All three are mission critical. Without the coordinator, workers cannot access the list of tasks, and no more work can be done. Without a configuration manager, many services are unable to function as they cannot learn about the other components in the system. And, if long-running compute jobs fail, they have to start from scratch which can slow down the progress of the system as a whole. CheckSync can transparently support all of these applications. Because CheckSync lives in the runtime, it cannot support all applications. This is discussed further in \cref{sec:appsupport}.

\subsection{Overview}
\label{sec:overview}
CheckSync's design is composed of five components: a configuration service, the two managers (\cref{sec:manager}), the checkpointer (\cref{sec:checkpointing}) and the restorer (\cref{sec:restore}). The configuration service is responsible for determining who the current primary is and directing clients to it, as well as managing failover. The managers are responsible for initiating the checkpointing and restoration processes and handling storage of checkpoints. The checkpoint/restore code is responsible for doing the work of checkpointing the application and restoring it to a running state from a checkpoint. CheckSync has two different operating modes, asynchronous and synchronous. Asynchronous replicates checkpoints in the background, allowing the application to resume as soon as the checkpoint is written to the primary's disk, while synchronous waits until replication completes to resume operation, in addition to checkpointing before replying to clients (\cref{sec:consistency}).

CheckSync relies on the language runtime to supply it with some specific information and functionality in order to simplify the checkpointing process and to deliver good performance. First, it uses the scheduler's knowledge of thread state to checkpoint only at safe points. It uses the runtime's knowledge of the application's memory to take incremental checkpoints. Third, it uses the runtime's knowledge of allocated data structures to enable synchronous replication. The implementation of CheckSync is integrated into the Go\cite{golang} runtime, which maintains the required information. We believe that Java and C\# also have runtimes that support these features.

\begin{figure}[t]
    \def\svgwidth{\hsize}
    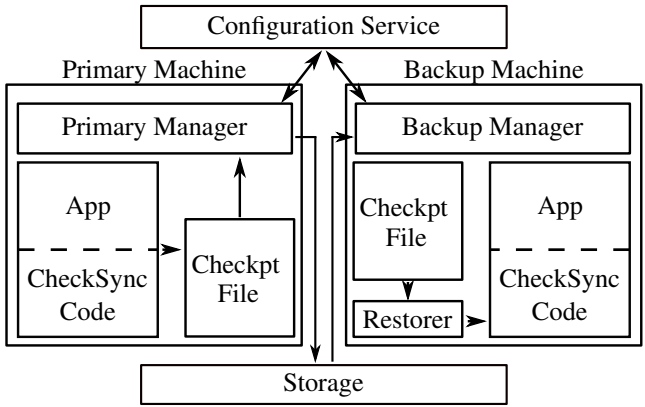
    \captionsetup{singlelinecheck=off}
    \caption[]{ CheckSync Workflow. The primary takes a checkpoint and sends the resulting file saved to storage. A configuration service detects when the primary fails and assigns a backup to take over when it does. The backup reads and restores from the stored checkpoint.}
    \label{fig:overview}
\end{figure}

\Cref{fig:overview}, illustrates the high-level workflow for CheckSync. The primary manager replicates the checkpoints so that they are available to the backup. A configuration service monitors the primary to detect failure, and assigns a backup up when it detects one. The backup reads and restores from the stored checkpoint.

\subsection{Managers}
\label{sec:manager}
CheckSync uses two managers to handle system operation, one on the primary and one on the backup. The primary manager has two roles: it starts the checkpoint process and replicates the checkpoints to a storage system. The backup manager is responsible for setting up and initiating the recovery process.

Users configure the checkpoint process through the primary manager by specifying paramaters like checkpoint frequency. The primary manager then starts the checkpoint process at this interval and stores the checkpoints produced. CheckSync relies on the existence of a distributed storage system for this, as that ensures that the checkpoints are themselves highly available. The primary manager copies the checkpoints to this storage system in the background. In addition to replicating the checkpoints, the primary manager also sends periodic heartbeat messages to a configuration service that is responsible for determining when a failure has occurred.

As discussed in \cref{sec:dump}, CheckSync takes incremental checkpoints. The primary sends each incremental checkpoint to storage and relies on the backup to do checkpoint reconstruction, as discussed in \cref{sec:restore}.

The backup manager is responsible for reassembling incremental checkpoints and initiating the recover process. When failure occurs, the backup manager fetches the checkpoints from storage and reassembles them into a complete checkpoint. It then triggers the restoration process, feeding it the complete checkpoint.

\subsection{Checkpointing}
\label{sec:checkpointing}
At a high level, the checkpoint process is divided into two phases.
\begin{enumerate}[label=\textbf{\arabic* -},itemsep=-1ex]
    \item \textbf{Suspension}. CheckSync takes care to suspend the running application in a state where no deadlocks can occur during the rest of the checkpointing process.
    \item \textbf{Dump}. CheckSync captures the state of the application and outputs it as a collection of files.
\end{enumerate}

\begin{figure}[h]
    \centering
    \def\svgwidth{\hsize}
    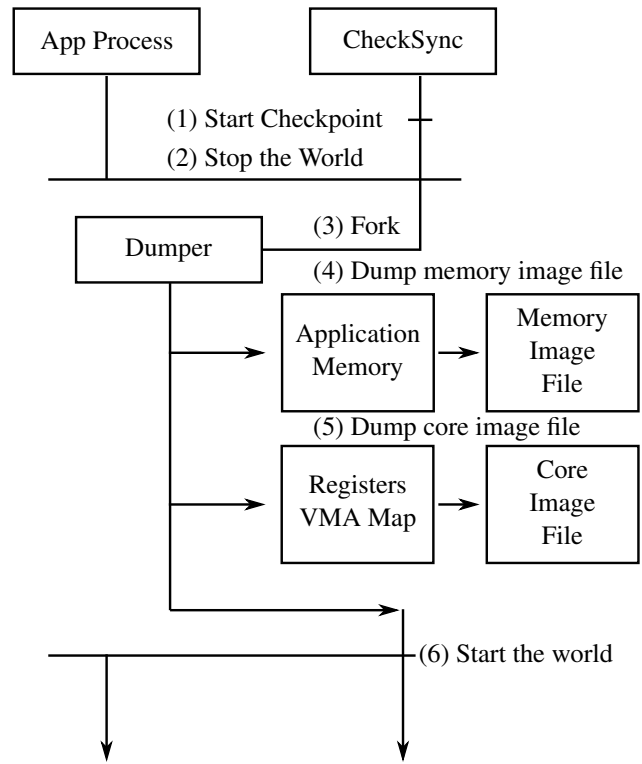
    \caption{The checkpoint operation. CheckSync runs in the background and periodically stops the world to checkpoint. After it finishes dumping the images, it starts the world again and allows the application to resume.}
    \label{fig:checkpoint}
\end{figure}

\subsubsection{Suspension}
Checkpointing occurs at an interval specified by the user when starting the primary manager. At each interval, CheckSync code running in the language runtime initiates the checkpointing process. In order to do so, CheckSync must be sure that the checkpoint it will take represents a consistent point in time. This requires that application execution is halted. Second, to avoid deadlock, the application must be halted in a state such that none of its threads are holding locks CheckSync will need.

CheckSync begins suspension by brining all running threads to a halt, aside from the one doing the checkpointing. CheckSync can do this because of it's integration with the language runtime, which schedules the execution of all the threads running in the application. CheckSync hooks into the scheduler and freezes each thread. However, freezing threads can have potentially harmful consequences. Consider the case where a thread is in the middle of allocating some memory. To do so, it must acquire a lock so that it can modify the allocator's free list, which it will be holding while it is frozen. However, CheckSync also needs that lock to safely access the information about what is free in the heap to facilitate incremental checkpointing. If CheckSync blindly froze the thread, it would therefore result in a deadlock. For this reason, CheckSync must run each task to a safe suspension point before suspending.

The conditions for safe suspension are a subset of the conditions needed for a garbage collector to suspend a running thread. CheckSync therefore leverages its integration with the language runtime to invoke the suspension step of the garbage collector and bring the application to a halt. While garbage collectors often have stricter conditions for suspending a thread than what CheckSync needs, such as needing to be certain that a thread is not manipulating hidden pointers, our evaluation shows that waiting to suspend does not significantly impact CheckSync's performance.

Once all the threads running in the application are suspended, all that is left are those associated with the runtime, including the checkpointing code. Therefore, it is safe for CheckSync to capture the memory of the application, as it knows nothing will modify it any further.

\subsubsection{Dump}
\label{sec:dump}
CheckSync must include all the state needed to reconstruct the application on the backup in the checkpoint. There are three things that it needs to dump to do this. First, the memory of the application. This contains all the state the application needs to execute. Second, the values of all the registers. This tells the resumed application where in memory to resume executing. Third, any open file descriptors and the files they refer to. The application may have open network sockets, epoll files, pipes, or on-disk files, and the resumed application will crash when it tries to read from them if they aren't restored correctly.

Capturing the memory of the application is done by reading all the virtual memory areas (VMAs) the process is using. CheckSync learns about this information by asking the operating system for it. On Linux, this is done using the \texttt{proc} pseudo-filesystem. CheckSync gathers this information while the application is suspended.

However, not all of an application's memory space will be modified between checkpoints. Were CheckSync to write out all of the application's memory each time, it would result in significant data duplication, and create bloated checkpoint sizes as they would grow linearly with the size of an application in memory. For applications like caches or lock servers this presents a performance bottleneck, which CheckSync avoids by using incremental checkpoints.

CheckSync uses a two-step process to determine what pages actually need to be dumped. First, it uses the operating system to find out which areas of memory have changed since the last checkpoint. On Linux this is done by checking the dirty bits in the \texttt{pagemap} file found under \texttt{/proc}. After taking a checkpoint, CheckSync resets this information so that the next read from it will contain only what has changed since the last checkpoint.

However, the first step may include pages that contain only dead objects in memory. CheckSync makes a second refinement pass over the set of remaining pages to avoid dumping such areas of memory.

Because the garbage collector and memory allocator inside the runtime already track information about areas of dead/alive memory, CheckSync queries them to find the set of pages that contain dead objects and then subtracts those from the set of pages returned in the first step. This allows CheckSync to construct a minimal set of pages that need to be checkpointed in a way that isn't replicable by techniques like virtual machine live migration that don't have access to the language runtime.

In addition to the memory of the application, CheckSync also needs to dump the critical state of the application such as registers and open file descriptors. CheckSync forks a child process to dump both this and the application's memory, the ``dumper''. It does this so that the dumper can inspect its parent to capture the parent's register values. This is necessary because thee parent cannot dump out its own register values, as this would add a function call to the stack which would invalidate the memory image captured earlier\footnote{This is due to a peculiarity of Go, which does not allow inlining of assembly without a dedicated function call}. The dumper dumps the contents of memory to the ``memory image file'', and the register values to the ``core image file''.

The core image file is also used to store metadata about the VMAs. In order to correctly reconstruct the process' memory state, CheckSync needs to capture the permissions for each VMA, as well as making note of VMAs that the process cannot read itself. These VMAs are primarily copy-on-write mappings and the pages of memory used for virtual system calls. Although the contents cannot be read and then written to the memory image file, their locations and sizes are included in the core image file as well so that the backup also has these mappings. This is important as otherwise the language's memory allocator could allocate those pages for the application to use, which would cause a crash if the copy-on-write areas need to be accessed later.

Also included in the checkpoint is information about open pipes, sockets, and epoll files. These are all important as many networking libraries use them, and CheckSync needs to be able to support those libraries for client/server applications to function. The contents of open pipes and sockets are included in the checkpoints as well, in their own files.

The parent of the dumper simply waits for the dumper to finish execution, at which point it starts the world again and reallocates the original number of threads to the application. This ensures that the application is never actually forced to run in a serialized form, as it is forced onto a single thread only while its execution is stopped.

\subsection{Restoration}
\label{sec:restore}
\begin{figure}[h]
    \centering
    \def\svgwidth{\hsize}
    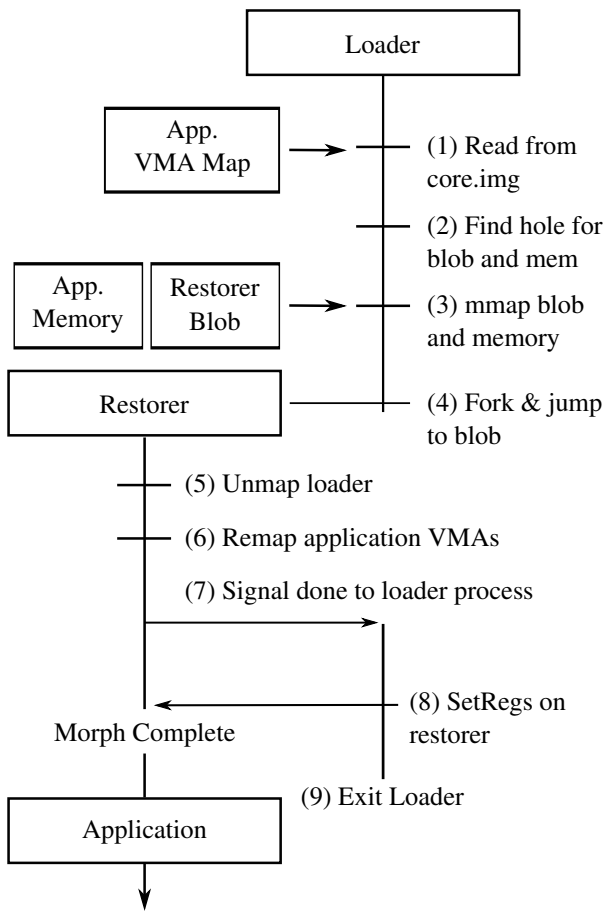
    \caption[Restore]{The CheckSync Restoration Process.}
    \label{fig:restore}
\end{figure}

Restoration is initiated by the backup manager when it detects that the primary has failed. When this happens, the backup manager begins by reconstructing the partial checkpoints made available by the primary, creating a complete checkpoint that it uses to do the restoration. The restoration process is then undertaken by carefully recreating the application's memory space and then jumping into it to resume execution.

\subsubsection{Checkpoint Reconstruction}
The first step in restoration is to build a complete checkpoint from the incremental checkpoints in storage. The backup manager fetches the incremental checkpoints when restoration is triggered by and then merges them into a complete checkpoint. Then, the backup starts restoring from that checkpoint. In order to avoid having the backup merge an unbounded number of incremental checkpoints, the configuration service periodically triggers a service that merges the incremental checkpoints in storage into a single checkpoint which can then be used as the starting point for future merges.

Merging checkpoints is simple, which we illustrate by describing a merge of two checkpoints. Merging more than two checkpoints is a repeated application of the following. We refer to checkpoint 1 as the earlier of the two in chronological order.
\begin{enumerate}[label=\arabic*)]
    \item The backup manager reads the core image file for each checkpoint, noting which pages of memory they touch
    \item The backup manager takes all VMAs from checkpoint 2 and overwrites the pages associated with those locations in checkpoint 1 with the data from checkpoint 2
    \item The backup manager modifies the metadata in the core image file of checkpoint 1 to reflect the updated register values and mappings from checkpoint 2
\end{enumerate}
After this, checkpoint 1 represents a complete merge of both checkpoints, and checkpoint 2 can be deleted.

\subsubsection{Load/Restore}
\begin{figure}[h]
    \centering
    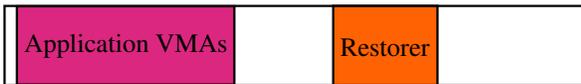
    \caption[Restore]{The progressive reconstruction of the application's memory space during the restoration process.}
    \label{fig:memlayout}
\end{figure}
Once CheckSync has a complete checkpoint it can start the restoration process. At the end of restoration, CheckSync must have constructed a process identical to the original application in memory layout, register values and open file descriptors. CheckSync does this by starting with a loader process that gradually morphs into the application process. This morph is necessary to ensure that all of memory has the correct permissions and behaves identically to how it did on the primary. Doing so is tricky, as the application's memory is complicated and contains multiple mmap'd areas separated by holes. CheckSync must recreate this layout exactly.

The loader begins by reading the core image file and looking at the table containing the information about the application's VMAs. It also parses the proc filesystem to determine its own VMAs. There is no guarantee that there is no overlap in these VMAs, so the loader relies on a separate component, the restorer, to handle possible conflicts. It is the loader's job to put all the contents of the memory image file into memory \emph{somewhere}, and then the restorer's job to move it to the correct locations and start executing. The process for doing so is illustrated in \cref{fig:memlayout}.

For this to be possible, the restorer is a PIE-compiled binary compiled using CRIU's compel utility\cite{Compel}. The result is a blob of executable code of a fixed size. The loader reads the size of the blob, finds a hole in memory where the blob will not overlap with any of the application's VMAs, and maps the blob there. Then, the loader maps the contents of the memory image file in a location that also doesn't overlap with the location of the application VMAs.

With this done, the loader process' memory contains: the restorer blob, the loader itself, and all the memory of the application. At this point the loader opens the files that were open on primary as well as creating sockets and internal pipes to match those that were open when the application was created. It uses the \texttt{dup} system call to move these into the proper file descriptors before continuing.

CheckSync does not need to perform any specific reconstruction of application threads. Instead, it relies on the runtime scheduler to recognize the threads in memory and to schedule them itself.

With all necessary information loaded into memory and thee descriptors also allocated correctly, the loader allocates a structure with all the metadata needed to restore the VMAs to their proper locations, and executes a jump instruction into the restorer with the address of this structure on the stack. The restorer can then read this structure, using the data it contains to map the application VMAs to their proper locations.

Once this is completed, the restorer executes a \texttt{ptrace} command to overwrite its registers with the values stored in the checkpoint, jumping execution into the application's virtual memory space and completing the restoration process.
\subsection{Synchronous CheckSync}
\label{sec:consistency}

Failures may occur between checkpoint intervals, which could cause clients to receive a success message for an operation on the primary, only for the changes made by that operation to disappear after failure because no checkpoint was taken that included it. This is fine for an application like the MapReduce coordinator, as it will result in some worker jobs being run more than once. While this is a bit wasteful, it doesn't impact the correctness of the application.

However, this is not true for all applications. To support them, CheckSync offers an option to do replication synchronously. Synchronous CheckSync relies on application developers to use a provided library function to mark places where state is exposed to clients, and takes a checkpoint at these points in the code. Synchronous CheckSync waits until the checkpoint is safely stored before resuming the application. This solution functions well for applications that encapsulates all their state in a single or small set of data structures, such as lock servers, but performs worse than asynchronous CheckSync due to the forced serialization is imposes on client requests.
\subsection{Failover}
\label{sec:failover}
The configuration manager is responsible for monitoring heartbeats from the primary, which it uses to detect failures. In order to make failover as seamless as possible, the application must be quickly resumed on a backup machine and then clients must be directed to that backup machine so that they can continue issuing their requests. Additionally, the application and clients may need to handle duplicate detection and retransmit earlier request to handle the consequences of a failover.

The primary manager periodically sends heartbeat messages to the configuration service. When enough of these messages are missed, the configuration service initiates failover. It selects a backup machine to use as the new primary, and initiates the restoration process on the backup. Once this is completed, the configuration service marks the backup as the new primary and directs all future client requests to that machine instead of the old primary.

Clients need to be prepared to deal with failover as well as the application being checkpointed. TCP connections are not carried over in a failure because they exist in the kernel, outside of the space of the application. This means clients will see a failure as a loss of connection. They must be prepared to handle this by reconnecting to the configuration service, while will eventually point them to the new primary.

A checkpoint followed by a failure may occur between a request finishing and a response being sent. Upon resuming, the application will try and send the response again, but because the TCP sockets are not copied over, this will fail. For this reason, the client may retransmit that request, which has already been processed, even with synchronous CheckSync. CheckSync does not automatically handle this, instead relying on the application to implement duplicate request detection if necessary. We consider this an area of future work.

Failover can create another anomalous situation for asynchronous CHeckSync. Consider a scenario where 2 update requests execute in parallel. On the primary, the first request finishes executing before the second, but reordering on the backup caused the second request to execute first there. Any client that read the value that was updated on the primary will expect it to be what was set by the second request, but on the backup the value is actually what was set by the first request. There are applications unaffected by this, such as the MapReduce coordinator which can safely use asynchronous CheckSync, or caches which care little for reordering. Other applications can mitigate this by using synchronous CheckSync.

Most clients already take care to automatically reconnect on a loss of connection, but some may need to be modified to include this. Additionally, application specific modifications may need to be made to the client code to handle request reordering and to support duplicate detection.

\subsection{Application Support}
\label{sec:appsupport}
Not all applications can use CheckSync. Because CheckSync lives inside the language runtime, it has no knowledge of the wider state of the system, or other processes running on the same machine. This prevents applications that make use of fork/exec from using CheckSync, as well as applications that use inter-process communication.

As noted earlier, CheckSync cannot replicate the TCP state stored in the kernel, and thus loses all active connections after a failover. This means that clients will have to reconnect after a failure. If an application relies on TCP status to perform other tasks, it will either need to be redesigned or is not compatible with CheckSync.

While CheckSync is able to replicate open files by tracking the applications active file descriptors and copying the files they refer to into storage alongside the checkpoint, it is not designed to replicate large amounts of data stored in files and performance will suffer significantly if large files are used. CheckSync's file replication is designed to support pipes, sockets, and epoll files more than it is to handle actual on-disk files.

We believe that many mission-critical applications do not rely on any of the above, and have provided three examples of such at the start of this section.

The features CheckSync doesn't support are summarized in \cref{tab:applimits}.

\begin{table}[h]
    \begin{center}
    \begin{tabular}{c | c}
        \textbf{Feature} & \textbf{Status}\\
        \hline\hline\\
        Multiple processes & Not supported \\
        IPC & Not supported \\
        Kernel state & Not replicated \\
        Large files & Slow performance \\
    \end{tabular}
    \end{center}
    \caption{Limitations on apps that use CheckSync}
    \label{tab:applimits}
\end{table}

\section{Implementation}
\label{sec:implementation}

\begin{figure}[h]
    \centering
    \def\svgwidth{\hsize}
    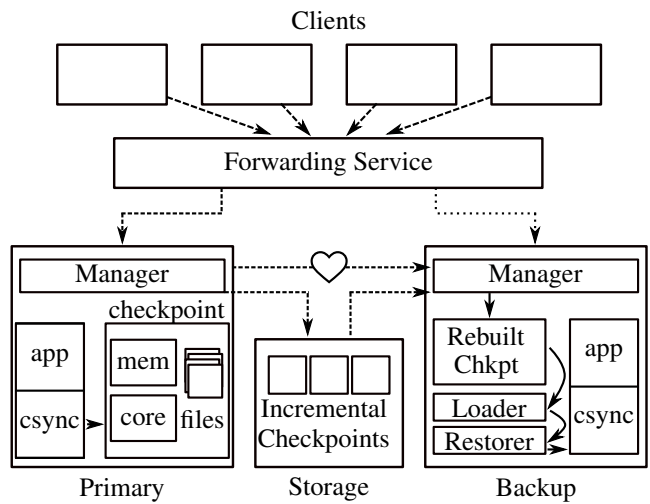
    \caption[Restore]{CheckSync as Implemented}
    \label{fig:deploy}
\end{figure}

We implemented CheckSync inside the Go language runtime. The checkpointing code and manager code were all written in Go, while the loader and restorer were both written in C. CheckSync relies on the Go runtime to monitor and freeze application threads at safe points, for its information about the liveness of areas of memory, and for its knowledge of where in memory objects are located. We believe that CheckSync could be implemented for any other language whose runtime provides these features.

\Cref{fig:deploy} illustrates the full implementation and functionality of CheckSync, which uses two additional machines. One is used to store the checkpoint files for fault tolerance and can be replaced with any fault tolerant storage service such as Amazon S3, and the second is a configuration manager. This machine is responsible for deciding which machine is the current primary and directing clients to that machine only and for initiating the failover process.

\begin{table}[h]
    \begin{center}
    \begin{tabular}{l | r}
        \textbf{Component} & \textbf{Lines of Code} \\
        \hline\hline\\
        Changes to Go runtime & $150$\\
        CheckSync Library Code & $600$\\
        Loader & $400$\\
        Restorer & $300$\\
        Manager Code & $500$\\
        Configuration Manager & $200$
    \end{tabular}
    \end{center}
    \caption{Lines of code for CheckSync's components}
    \label{tab:loc}
\end{table}
\Cref{tab:loc} lists the components of CheckSync and the lines of code used to implement each of them. CheckSync's modifications to the Go runtime are kept small by relying on a Go library to do most of the work. The only changes needed in the runtime itself were to expose necessary functions and information to the CheckSync library.

The primary and backup managers communicate with the configuration manager using gRPC\cite{grpc}.

\section{Evaluation}
\label{sec:eval}

CheckSync is intended to be nearly as transparent to applications as
virtual machine checkpointing, but with higher performance. This
section evaluates both of these properties by answering four
questions:

\begin{itemize}
\item How much effort is required to adapt three existing applications
  to use CheckSync? (\ref{sec:ease});

\item What is the impact of checkpointing on normal-case application
  throughput? (\ref{sec:perfeval});

\item How large are CheckSync's checkpoints? (\ref{sec:size});

\item How long does it take to recover from a failure? (\ref{sec:deployment})

\end{itemize}

\textbf{Deployment.} We deployed CheckSync on a Cloudlab deployment using c220g2 machines. These machines each have 2 Intel 25-2660 CPUs with 10 cores each running at 2.60GHz, 160GB of RAM and 10Gb NICs. Two machines were used as the primary and backup, while a third and fourth machines were used as the configuration manager and storage service respectively. All machines were co-located in the Cloudlab Wisconsin data center.

\subsection{Ease of use}
\label{sec:ease}

To evaluate CheckSync's ease of use, we selected three existing applications and ran them with CheckSync. For each application, we measured the lines of code needed to get it to operate on top of CheckSync, and also conducted a basic test to ensure that the application operated as expected.

\textbf{Application Selection}.
We selected three existing applications to use with CheckSync, each of which has different characteristics. First, we took an implementation of the MapReduce coordinator written previously. Second, the \texttt{go-cache} key/value store. And third, a benchmark for the gonum package.

All three of these applications adhere to the restrictions described in \cref{sec:appsupport}. Also, all three of these applications have different characteristics. The MapReduce coordinator represents an ideal mission-critical application which requires communication with multiple workers over the network and makes extensive use of parallelism.

The \texttt{go-cache} key/value store also requires network connections, but places stresses RAM more than the coordinator. We evaluate the performance of this application in \cref{sec:perfeval}. Also, many mission-critical applications can be abstractly considered key/value stores. Lock servers and caches, for example, both requiring maintaining structures like a key/value store in order to function.

We also evaluate a benchmark for the gonum package as that places a much higher emphasis on compute resources than the other applications. This benchmark is heavily parallel, and stresses the CPU by performing many optimization operations in a row.

\begin{table}
    \begin{tabular}[t]{ l | r | r }
        \textbf{Application} & \textbf{CheckSync} & \textbf{VM} \\
        \hline\hline
        MapReduce Coordinator & 0 & 0\\
        MapReduce Client & 5 & 0 \\
        \texttt{go-cache} and client & 0 & 0\\
        gonum & 0 & 0
    \end{tabular}
    \caption{Lines of code changed to get three applications working on CheckSync and a virtual machine using Remus}\label{tab:locchanged}
\end{table}

\textbf{Results.} \Cref{tab:locchanged} shows the lines of code that had to be changed to get each of the applications working on CheckSync as compared to a virtual machine.

Both CheckSync and the virtual machine approach required no changes to the application code for gonum and \texttt{go-cache}, and no changes to the \texttt{go-cache} client\footnote{We used the YCSB client, which already tries to reconnect if a failure happens}. While CheckSync didn't require any changes to the coordinator either, the worker code did have to be modified to retry connecting to the coordinator if it could not be reached due to an in-progress failover.

This demonstrates that CheckSync is as easy-to-use as Remus virtual machine migration for the applications that it supports.

\subsection{Throughput Overhead}
\label{sec:perfeval}
This section evaluates the decrease in \texttt{go-cache} throughput due to CheckSync. We evaluate both asynchronous and synchronous CheckSync. We compare this overhead to that imposed by using virtual machines running on the Xen hypervisor and process checkpointing using CRIU\cite{noauthor_checkpoint-restorecriu_2021}. The virtual machines were allocated by CloudLab with 12 cores (the max allowed by CloudLab) and 64GB of RAM. We restricted \texttt{go-cache} to use only 12 cores in all our evaluations to ensure fairness of the comparison.

Unfortunately, we were unable to get Remus working on CloudLab due to problems with networking when installing the Xen hypervisor on bare metal machines. However, we were able to get a local version of Remus working to evaluate the overhead it introduced on top of the Xen virtual machines.

Additionally, we compare the overhead introduced by CheckSync against the overhead introduced by \texttt{go-cache}'s snapshot system. An application specific snapshotting scheme like \texttt{go-cache}'s writes only the data inside the hash table that stores values to disk.

\textbf{Benchmark.} We use YCSB\cite{cooper_benchmarking_2010} to generate two workloads that we benchmark against. YCSB has been widely used to evaluate the performance of production systems. Both workloads run a million operations against 1000 keys. Workload A is 50\% updates, 50\% gets. Workoad B is 5\% updates, 95\% gets. Each entry is 1000 bytes long.

A single client was used to run the YCSB workload, which used 20 worker threads to place parallel load on the server. Any more than 20 worker threads resulted in no measurable increase in throughput. We implemented a \texttt{go-cache} backend for YCSB, and use that for both the virtual machine approach and CheckSync.

\textbf{Configuration.} CheckSync, Remus and CRIU were all configured to checkpoint/migrate every 200ms, the suggested migration frequency for Remus. \texttt{go-cache} is snapshotted every 200ms as well. Checkpoints are saved to disk, as are snapshots. To account for this in our Remus evaluations, we had Remus perform migration to the same machine.

The exception to this is for synchronous CheckSync, which replicates its checkpoints to a machine in the same data center. These machines share a 10Gb ethernet link and have a latency of \~ 150 $\mu s$ between them.

\begin{figure}[t]
    \scalebox{1.0}{\input{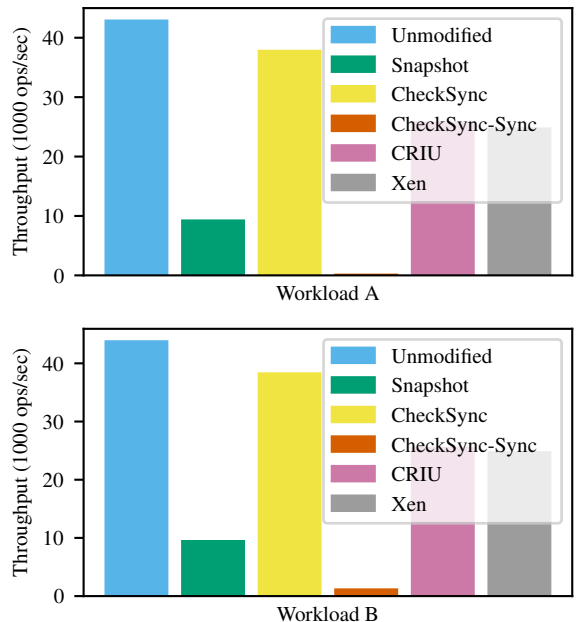}}
    \caption[]{Throughput of \texttt{go-cache} in two different workloads and five different configurations: \texttt{go-cache}, \texttt{go-cache} with snapshotting enabled, \texttt{go-cache} on CheckSync, \texttt{go-cache} on CheckSync-Sync, and \texttt{go-cache} on a Xen virtual machine.}
    \label{fig:workloadatp}
\end{figure}

\begin{table}
    \centering
    \begin{tabular}[h]{l | r | r}
        \textbf{System} & \textbf{Workload A} & \textbf{Workload B} \\
        \hline\hline
        CheckSync & 11.88\% & 12.61\% \\
        CheckSyncSync & 99.83\% & 97.48\% \\
        GoCache w/ Snapshots & 78.54\% & 78.48\% \\
        GoCache on Xen & 42.41\% & 43.59\% \\
        Est. GoCache w/ Remus & 68.32\% & 65.02\% \\
        GoCache w/ CRIU & 45.58\% & 46.80\% \\
    \end{tabular}
    \caption{The drop in throughput compared to running an application on its own for CheckSync and baselines}\label{tab:overhead}
\end{table}
\textbf{Results.} \Cref{fig:workloadatp} shows the results of running \texttt{go-cache} on its own, on top of CheckSync with asynchronous replication, CheckSync with synchronous replication, with CRIU checkpointing it, and inside a virtual machine. Asynchronous CheckSync outperforms the all the other configurations, with an overhead of only 12\%. We omit the Remus measurements from this graph as they use estimates made on a different setup.

The 12\% overhead introduced by CheckSync is due to the time an application spends suspended while checkpointing. Most of this time (65\%) is spent interacting with the file system while CheckSync reads from \texttt{/proc} and writes the checkpoint to disk. The remaining time is spent managing the threads and doing the computations that facilitate incremental checkpointing.

\Cref{tab:overhead} shows the percentage of throughput lost when using \texttt{go-cache} in different configurations. Each line in the table shows the overhead introduced by a strategy, with a 10\% overhead representing a 10\% loss in throughput compared to the baseline. All \texttt{go-cache} deployments use \texttt{go-cache} on a c220g machine as a baseline. Our local measurement of Remus had an added overhead of 50-60\% on top of a Xen virtual machine, which matches the numbers from the Remus paper. We combine the local overhead for each workload with the overhead of Xen in CloudLab to produce the estimate tabulated in the last row of \cref{tab:overhead}.

The results show that asynchronous CheckSync incurs a relatively small overhead when compared to \texttt{go-cache} snapshotting, CRIU and Xen/Remus. The gap between CheckSync and Remus is particularly large, confirming our hypothesis that virtual machines add an unnecessary amount of overhead to small applications like the one's we tested.

Additionally, the numbers in \cref{tab:overhead} show that CheckSync with asynchronous replication outperforms CRIU, which has a 45\% throughput overhead. This is because CRIU's checkpoints are larger as it has to checkpoint more of the operating system state that CheckSync.

Interestingly, as we show in \cref{tab:checkpointsize}, \texttt{go-cache} snapshots are smaller than CheckSync checkpoints, but still add more overhead. We believe this to be for two reasons. First, \texttt{go-cache} uses \texttt{gob}\cite{go-gob} to serialize the hash table. This is more expensive than CheckSync, as it requires that Go fully traverse the map, following all pointers it contains. CheckSync, however, just dumps pages of memory directly and doesn't need to traverse anything. Additionally, snapshotting in \texttt{go-cache} allows all other threads to continue executing, but does hold a lock on the hash table. This means that all threads will eventually get stuck waiting for the lock, and then will execute serially after the snapshot is dumped until there is no more contention for the lock. CheckSync avoids this by pausing all threads while the checkpoint is taken, which means there is no extra lock contention after the checkpoint finishes.

Synchronous CheckSync doesn't perform nearly as well as asynchronous CheckSync does. This is expected as it checkpoints more frequently and forces a serialization of requests that reduces performance due to a loss of parallelism. Both of the other comparison points are also periodic, and snapshot far less often than synchronous CheckSync does. We discuss potential optimizations to synchronous checkpointing in \cref{sec:futurework}.

Our results show that CheckSync reduces throughput by 12\%, which is better than the reduction experienced when using CRIU, snapshots and Remus.
\subsection{Checkpoint Size}
\label{sec:size}
\begin{table}[b]
    \centering
    \begin{tabular}{l | r}
        \textbf{System} & \textbf{Size (KB)} \\
        \hline\hline
        CheckSync & 1229\\
        Virtual Machine & 889000\\
        \texttt{go-cache} snapshots & 738 \\
        CRIU incremental checkpoints & 3794
    \end{tabular}
    \caption{The average size in KB of the checkpoints/snapshots written to disk by CheckSync and other systems over the course of a run of Workload A}\label{tab:checkpointsize}
\end{table}

We showed in the previous section that interacting with the disk is the largest bottleneck to CheckSync's performance. For this reason, it is important that checkpoint size be as small as possible in order to reduce the amount of data that has to be written to disk.

We evaluate the size of CheckSync's checkpoints on a run of Workload A, during which the client issued 500,000 updates on 1000-byte values. The results of this comparison are tabulated in \cref{tab:checkpointsize}. As CheckSync with synchronous replication uses the same checkpointing technique as asynchronous CheckSync, the CheckSync data point captures the size of both schemes.

Virtual machine snapshots are by far the largest of the four comparison points, which makes sense as they include the most state, even if they are taken incrementally. CheckSync's checkpoints are an order of magnitude smaller than the virtual machine snapshots.

The reason for CheckSync's small size is due to the effectiveness of incremental checkpointing. Each checkpoint only contains the values that have been changed since the last checkpoint, which is why the checkpoints are smaller than the total size of the application.

CRIU checkpoints also use their own version of incremental checkpointing, which doesn't have access to the runtime information CheckSync does. Additionally, CRIU checkpoints more of the operating system. The combination of these two factors explains why CRIU checkpoints are three times larger than CheckSync's.

\texttt{go-cache} snapshots contain only the data from the hash table, and include none of the data from outside the application heap that CheckSync has to capture. This explains why the checkpoints are so small, which comes at the cost of a more complex resume process, as the application has to start from scratch unlike CheckSync. That CheckSync checkpoints are close to the size of these application-specific snapshots is a testament to incremental checkpointing, which we evaluate next.

\begin{table}
    \centering
    \begin{tabular}{c | c | c | c}
        \textbf{Workload} & \textbf{Initial} & \textbf{First Pass} & \textbf{Second Pass} \\
        \hline\hline
        Workload A & 419058.06 & 375.89 & 307.23 \\
        Workload B & 425603.02 & 437.83 & 340.95 \\
        Workload C & 440274.68 & 1328.88 & 882.16
    \end{tabular}
    \caption{The number of pages identified as having to be checkpointed at each step of the incremental checkpointing process. CheckSync only checkpoints the pages left after the second pass.}\label{tab:incrementalreduction}
\end{table}

\textbf{Impact of Incremental Checkpointing}. \Cref{tab:incrementalreduction} shows the impact of CheckSync's incremental checkpointing on checkpoint size. We constructed a separate, update dominated workload C to simulate the workloads imposed on services like lock servers, where each request eventually results in an update.

In all cases, the largest reduction in number of pages to be dumped came from the first pass. This makes sense, as the \texttt{/proc} pass has information about the process' memory that the runtime doesn't know about. The runtime only tells CheckSync information about the memory on the heap. This is most impactful when objects are frequently being garbage collected, which occurs most frequently when items are deleted from storage. This happens most frequently in Workload C, which explains the increased impact of the second pass in that case.

The results of our evaluation of checkpoint size show that CheckSync's incremental checkpointing successfully reduces checkpoint size by a significant amount. It is because CheckSync is able to keep checkpoint size small that it is able to achieve the throughput shown in \cref{sec:perfeval}.

\subsection{Deployment and Failover}
\label{sec:deployment}
In order to measure the time on CheckSync, we setup a machine to run \texttt{go-cache} and loaded it with 1000 keys, sending heartbeats to the configuration service once every 100ms. Then, we introduced a failure on the primary machine and waited for the application to resume on the backup and read back the 1000 keys to ensure the system operated correctly. We measured the time to recover from failure, which took 829ms, most of which was spent reconstructing and reading the checkpoints into memory. This means the applications are offline for less than a second, making it highly available.

The time to recover is proportional to the size of the application in memory, so applications that use more memory take more time to recover. This is because the bottleneck on recovery speed is mapping the contents of \texttt{mem.img} into memory. The larger \texttt{mem.img} is, the longer this takes, and the size of \texttt{mem.img} after merging all incremental checkpoints is identical to the size of the application in memory.

\section{Future Work}
\label{sec:futurework}
While CheckSync provides efficient, easy-to-use high availability for mission-critical applications, there are three directions for future work. First, improving the performance of CheckSync and synchronous CheckSync. Second, improving client/server support. Third, using CheckSync in serverless computing.

\textbf{Improved Performance.} While CheckSync produces small checkpoints, they are still larger than the snapshots produced by \texttt{go-cache}. This means there is more room for optimization in the checkpointing process. One way CheckSync could further reduce performance is by using the garbage collector to do a trace of all the live \emph{objects} in memory during the second pass of incremental checkpointing. Currently, if a page contains a live object, the whole page will be dumped. However, this could be reduced by dumping only the live object, which can be identified by reusing the pointer tracing the garbage collector already does.

Additionally, synchronous checkpointing is currently extremely slow compared to asynchronous checkpointing. While this is expected given the frequency at which it checkpoints, there are potential optimizations to improve its performance as well. For example, requests could be batched by the server before returning to the client. This way CheckSync can reduce the 1:1 ratio of state-modifying requests to checkpoint operations. This could significantly increase synchronous CheckSync's performance.

\textbf{Network Replication.} One of CheckSync's weaknesses is that clients have to reconnect to a server after a failover happens. This increases the time it takes to recover from failure and interrupts a seamless client experience. Virtual machine migration mitigates this by replicating the entire TCP stack from within the kernel, maintaining all existing connections. While CheckSync wouldn't be able to do this from within the runtime, it might be possible to extend the manager to do this instead. This would increase CheckSync's usability for client/server applications.

\textbf{Serverless Computing.} CheckSync has applications outside of providing high availability. Serverless computing is a growing field, and the complexity of the functions being run as services is steadily growing. Checkpoints are already used to enable hot-starts of serverless functions\cite{zhang_kappa_2020}, but these checkpoints do not support multithreaded code. CheckSync could be used to support hot-start of more complicated serverless functions, and also to provide fault-tolerance for serverless functions that run for an extended period of time.
\section{Conclusion}

CheckSync is a system designed to provide high availability to mission-critical applications. It takes an existing application written in a memory-managed language and periodically checkpoints and replicates the application's state which it can then resume from in the case of failure. CheckSync provides developers with the option of using asynchronous or synchronous checkpointing based on the needs of their application. Our evaluation demonstrates that CheckSync is as easy-to-use as a virtual machine for the applications it supports, while providing higher performance.

CheckSync represents a new option for ensuring the availability of mission-critical applications, thereby increasing the reliability of large-scale distributed systems.
\ifhidden
\section*{Acknowledgements}
Some acknowledgements should probably go here?
\fi
\iftemplate
\bibliographystyle{plain}
\bibliography{checksync}
\fi
\end{document}